%% file: main.tex
\documentclass[10pt, conference, compsocconf]{IEEEtran}
\usepackage{amsmath}
\usepackage{amsthm}
\usepackage{amssymb}
\usepackage{cite}
\usepackage{algorithm}
\usepackage{booktabs}
\usepackage{verbatim} 
\usepackage{subfigure}
\usepackage{balance}
\usepackage{graphicx}
\usepackage{tikz}
\usepackage{algpseudocode}
\usepackage[bottom]{footmisc}
\usetikzlibrary{fit,positioning}
\usepackage[utf8]{inputenc}
\usepackage[english]{babel}
\usepackage{enumitem} 
\usepackage{url}
\usepackage{cite}

\IEEEoverridecommandlockouts
\begin{document}

\title{Taureau: A Stock Market Movement Inference Framework Based on Twitter Sentiment Analysis}
	\author{\IEEEauthorblockN{Nicholas Milikich and Joshua Johnson}
	
	\IEEEauthorblockA{Department of Computer Science and Engineering\\
			University of Notre Dame\\
			Notre Dame, IN, USA\\
			\{nmilikic and jjohns54\}@nd.edu}
	}
	\maketitle

	\IEEEpeerreviewmaketitle
	\input{abstract}
	\input{introduction}

	\input{review}
	\input{problem}
	\input{proposed}

	\input{results}
	\input{further}
	
	\section*{Github Repository Link}	
    The Github repository for our code is available at the following link: \textbf{https://github.com/tahmidrashid/taureau}
	
	\bibliographystyle{IEEEtran}
% \clearpage%Page break. Remove before submission
\bibliography{refs} 

\end{document}

%% file: abstract.tex
\begin{abstract}
With the advent of fast paced information dissemination and retrieval, it has become inherently important to resort to automated means of predicting stock market prices. In this paper, we propose Taureau, a framework that leverages Twitter sentiment analysis for predicting stock market movement. The aim of our research is to determine whether Twitter, which is assumed to be representative of the general public, can give insight into the public perception of a particular company and has any correlation to that company's stock price movement. We intend to utilize this correlation to predict stock price movement. We first utilize Tweepy and getOldTweets to obtain historical tweets indicating public opinions for a set of top companies during periods of major events. We filter and label the tweets using standard programming libraries. We then vectorize and generate word embedding from the obtained tweets. Afterwards, we leverage TextBlob, a state-of-the-art sentiment analytics engine to assess and quantify the mood of the users based on the tweets. Next, we correlate the temporal dimensions of the obtained sentiment scores with monthly stock price movement data. Finally, we design and evaluate a predictive model to forecast stock price movement from lagged sentiment scores. We evaluate our framework using actual stock price movement data to assess its ability to predict movement direction.
\end{abstract}

\begin{IEEEkeywords}
stock market, Twitter, DJIA, Tweepy, getOldTweets, sentiment analysis, Word2Vec, TextBlob
\end{IEEEkeywords}

%% file: introduction.tex
\section{Introduction} \label{sec:introduction}
Due to the emergence of smart devices and widespread connectivity, \emph{social sensing} is progressing as a new sensing paradigm that is able to obtain real-time measurements about the physical world using people empowered with a plethora of devices~\cite{zhang2020transres}. Upon close examination of social media, we find interesting links between real-world occurrences and posts from the general public. This broad category of studying user behavior in social sensing is referred to as \textit{sentiment analysis}~\cite{zhang2020pqa}, where people's sentiments towards particular matters are observed and analyzed to infer important metrics. 

Predicting stock market prices is of growing importance both in academia as well as in industry. While stock price predictions often rely on historical data and macroeconomic trends, it has been shown that the general public's attitude in the form of Twitter sentiment is predictive of DJIA movement~\cite{bollen2011twitter}. Recent works have shown that such sentiment analysis can be applied in more specific cases. For instance, financial message boards can be analyzed by combined topic modeling and sentiment analysis~\cite{nguyen2015sentiment}, and industry-specific sentiment analysis dictionaries can be utilized to predict the stock price movement of specific companies~\cite{shah2018predicting}. However, these message boards are frequented by a small, specialized segment of the population. In this paper, we investigate the complex relationship between tweet sentiments (e.g., public opinions about particular companies) the financial market instruments (e.g., volatility, trading volume, and stock prices). We train a sentiment analysis tool that can use Twitter text to give a useful gauge of company performance, investigate a correlation between stock price movement and Twitter sentiments, and implement a forecasting model that can use these sentiments to predict future stock performance.

\begin{figure}[!h]
    \centering
    %\vspace{-0.08in}
    \includegraphics[width=8.5cm]{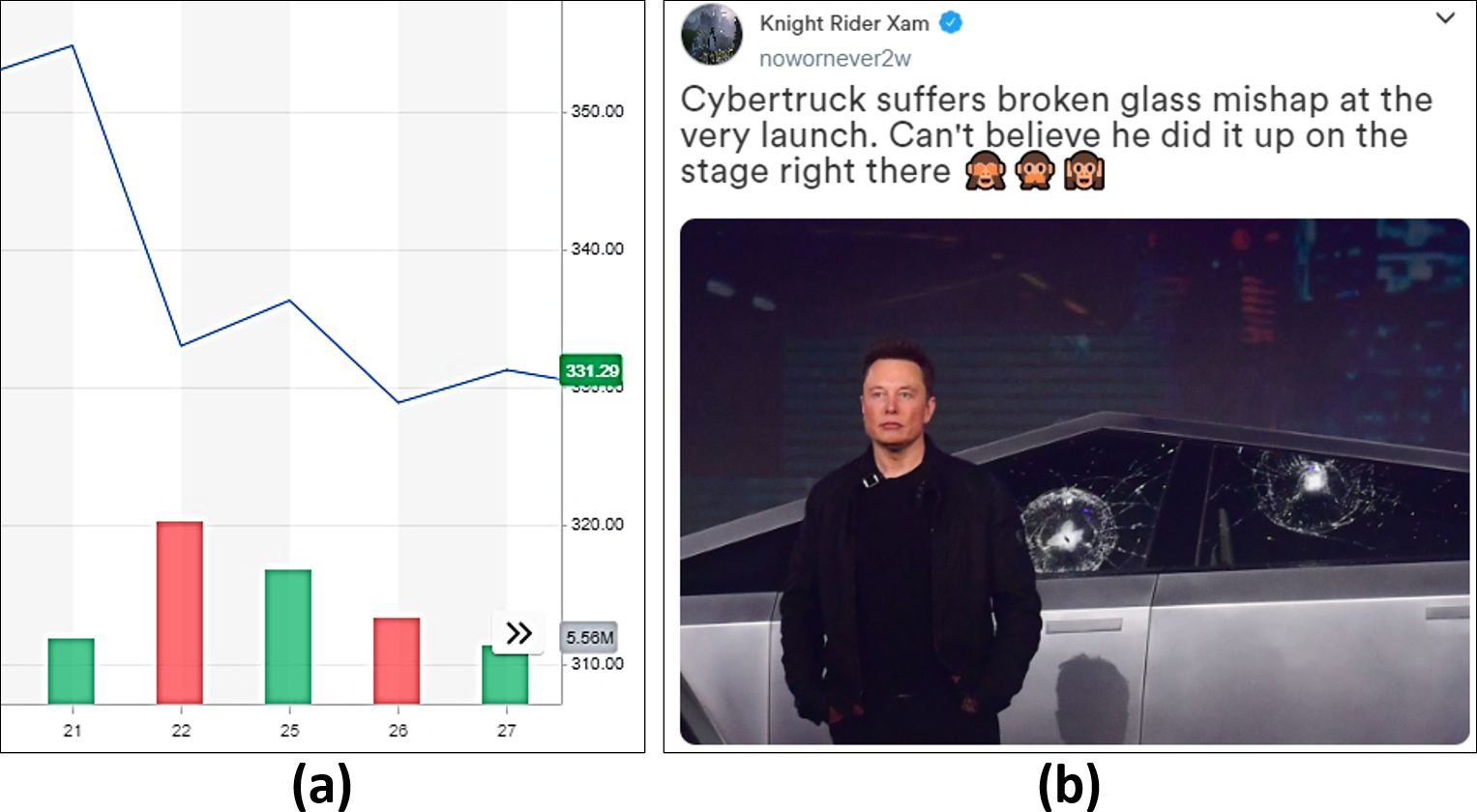}
    \vspace{-0.08in}
    \caption{(a) Stock prices of Tesla between November 21, 2019 and November 27, 2019; (b) A tweet posted on November 22 indicating the launch failure of Tesla}
    %\vspace{-0.08in}
    \label{fig:Tweets}
    %\vspace{-0.15in}
\end{figure}

Let us consider an incident that took place on November 21, 2019, when Tesla unveiled their new vehicle called the Cybertruck. During the Cybertruck's durability demonstration on stage, Elon Musk, Tesla's founder, unexpectedly smashed the windows of the vehicle using a metal ball. Figure~\ref{fig:Tweets} (a) shows the stock prices of Tesla between November 21, 2019 and November 27, 2019; and figure~\ref{fig:Tweets} (b) shows a tweet posted on November 22, 2019 that laments the launch failure of the Cybertruck. If we observe closely, we see that the stock price of Tesla on the same day (i.e., November 22, 2019) fell sharply. This implies that public perception of a particular company embedded in tweets is strongly correlated to the actual stock price movement of that company.

Intuitively, if the public opinion embedded into the tweets could be leveraged to predict the stock market price prediction, it could potentially help make more informed trade decisions for trading companies as well as individuals. However, a substantial amount of work is needed to transform the noisy and unreliable twitter data into a form that can be seamlessly analyzed and presented to the end users. With this basis, we proposes Taureau, a stock market movement inference framework based on Twitter sentiment analysis. The Taureau framework is able to collect tweets that indicate various opinions regarding several companies. The framework analyzes the tweets to obtain the sentiment of the users towards the companies using the TextBlob sentiment analysis engine~\cite{loria2018textblob} that employs an ensemble of different statistical approaches to generate the polarity and subjectivity of tweets. Based on the sentiments, our proposed framework is able to both predict the probable stock prices and recommend best practices to hold, buy, or sell the stocks.

%% file: review.tex
\section{Related Works} \label{sec:review}
In recent times, social sensing has received a significant amount of attention because of the widespread adoption of smart devices, low-cost mobile sensors~\cite{hasan2016development}, and the ubiquity of Internet connectivity~\cite{wang2018age}. A rich set of literature exists on social sensing applications ranging from intelligent transportation systems~\cite{liu2018urban}, and environmental sensing~\cite{mao2012citysee} to disaster and emergency response~\cite{rashid2019socialcar,rashid2020dasc}. Wang \textit{et al.}~\cite{wang2018age} provided a comprehensive study of state-of-the-art social sensing tools. 

With the onset of modern machine learning techniques, sentiment analysis has also garnered quite a good amount of attention in the field of Natural Language Processing (NLP). Several recent studies have leveraged sentiment analysis to predict spikes in book sales based on blog posts~\cite{gruhl2005predictive}, box office performance based on social media posts~\cite{apala2013prediction}, and inferring stock market performance based on Twitter data~\cite{mittal2012stock}. Medhat \textit{et al.}~\cite{medhat2014sentiment} carried out an extensive survey of contemporary sentiment analysis applications.

Recent studies have explored the relation between Twitter mood and DJIA movement. Bollen et al.~\cite{bollen2011twitter} extracted public mood from Twitter using two sentiment analysis techniques: OpinionFinder, which analyzes the text content of tweets to assign public mood a score on a one-dimensional positive-negative scale, and Google-Profile of Mood States, which assigns public mood a six-dimensional score characterizing the dimensions calm, alert, sure, vital, kind, and happy. The paper shows that certain dimensions of public mood, particularly calm and happy, improve standard stock market prediction models, while others, particularly OpinionFinder general happiness, do not. The paper validated the ability of these models to capture public opinion by cross-validating them over a time period including the 2008 presidential election and Thanksgiving day; the results were consistent with the expected public mood during these events. 

Nguyen et al.~\cite{nguyen2015sentiment} showed the effectiveness of using combined topic and sentiment modeling of Yahoo Finance message boards for 18 specific companies to predict stock price movement of those companies. Several studies have presented that it is possible to predict, to some degree of accuracy, stock price movement by textual analysis of public sentiment~\cite{rao2012analyzing,shah2018predicting,nguyen2015sentiment}. However, previous methods have gathered information from sources frequented by industry experts, a limited, specialized portion of the population. By contrast, the Taureau framework aims to determine whether more widespread public sentiment, in the form of Twitter data, is predictive of stock price movement for specific companies.

%Building on these techniques, other studies have demonstrated that such sentiment analysis techniques can be applied in more specific cases. For example,

%They specifically avoided using Twitter as a source for public opinions since tweets tend to be messier than message board posts, and since there is no consistent way to filter tweets by topic; this high level of noise makes it difficult to correlate posts to specific stocks. Further, Shah et al.~\cite{shah2018predicting} demonstrated the effectiveness of developing a sentiment analysis dictionary specific to financial markets (and to the pharmaceutical companies analyzed) in predicting directional stock price movement. %The authors observed that analyzing bigrams and trigrams was necessary to capture the meaning of the articles; for example, “declined” has a negative connotation, but “costs declined” has a positive connotation. However, they too analyzed public sentiment from a news outlet providing stock-specific news articles. 

%% file: problem.tex
%\newcommand{\me}{\mathrm{e}}
\newtheorem{myDef}{DEFINITION}
\section{Problem Definition}
In this section, we formally define the problem of our framework where we intend to accurately analyze social media data in order to predict the stock market movement. Our goal is to minimize the discrepancy between our predicted stock movement with the actual stock movement. We first define a few key definitions in our problem formulation and finally present our objective function. 

\begin{myDef}
\label{def:SD1}
\textbf{Social Sensing Data ($S$)}: We define the Social Sensing Data ($S$) to be the public opinion about companies posted on Twitter (e.g., tweets anticipating a company's new product, a potential merger, or a new of acquisition).
\end{myDef}

\begin{myDef}
\label{def:SD2}
\textbf{Companies ($\mathcal{L}$)}: We define a set of $L$ distinct companies $\mathcal{L}$. Each company is represented by index $l$ where $0<l\leq L$.
\end{myDef}

\begin{myDef}
\label{def:SD3}
\textbf{Predicted Stock Movement ($E_l$)}: We define the output for the \textit{predicted stock movement} of the companies using $E_l$ where $0\leq E_l\leq 1$. Intuitively, a higher value of $E_l$ indicates that company $l$ is performing well and vice versa for a lower value of $E_l$.
\end{myDef}

\begin{myDef}
\label{def:SD4}
\textbf{Ground Truth ($\widehat{E_l}$)}: We define the ground truth for the \textit{relative financial performance} of a company using $\widehat{E_l}$ where $0\leq \widehat{E_l}\leq 1$.
\end{myDef}

Using the above definitions, we define the goal of Taureau framework which is essentially a \textit{prediction} problem. Given the social media data inputs $\mathcal{S}$, the objective of the framework is to minimize the discrepancy between the predicted stock movement and their ground truths by solving a problem as follows:
\begin{equation}\label{eq:deadlsine}
 \begin{split}
 & \text{minimize:} \quad \sum_{l=1}^{L} (RMSE(\widehat{E_{l}},E_{l}) |  \mathcal{S})\\
% & \text{(drones and event deadlines constraint)}
\end{split}
\end{equation}\noindent
where $RMSE$ is a function to generate the root mean square error of a set of number. Root mean square error is essentially a standard deviation of the residuals (prediction errors).

%% file: proposed.tex
\section{Proposed Solution} \label{sec:proposed}
In this section, we propose the Taureau framework to leverage Twitter sentiment analysis for predicting stock market movement. Figure~\ref{fig:block} provides an overview of our framework. A series of steps are required to collect, analyze, and make the final predictions. The details of the steps are discussed below:

\begin{figure}[!htb]
    \centering
    \vspace{-0.08in}
    \includegraphics[width=8cm]{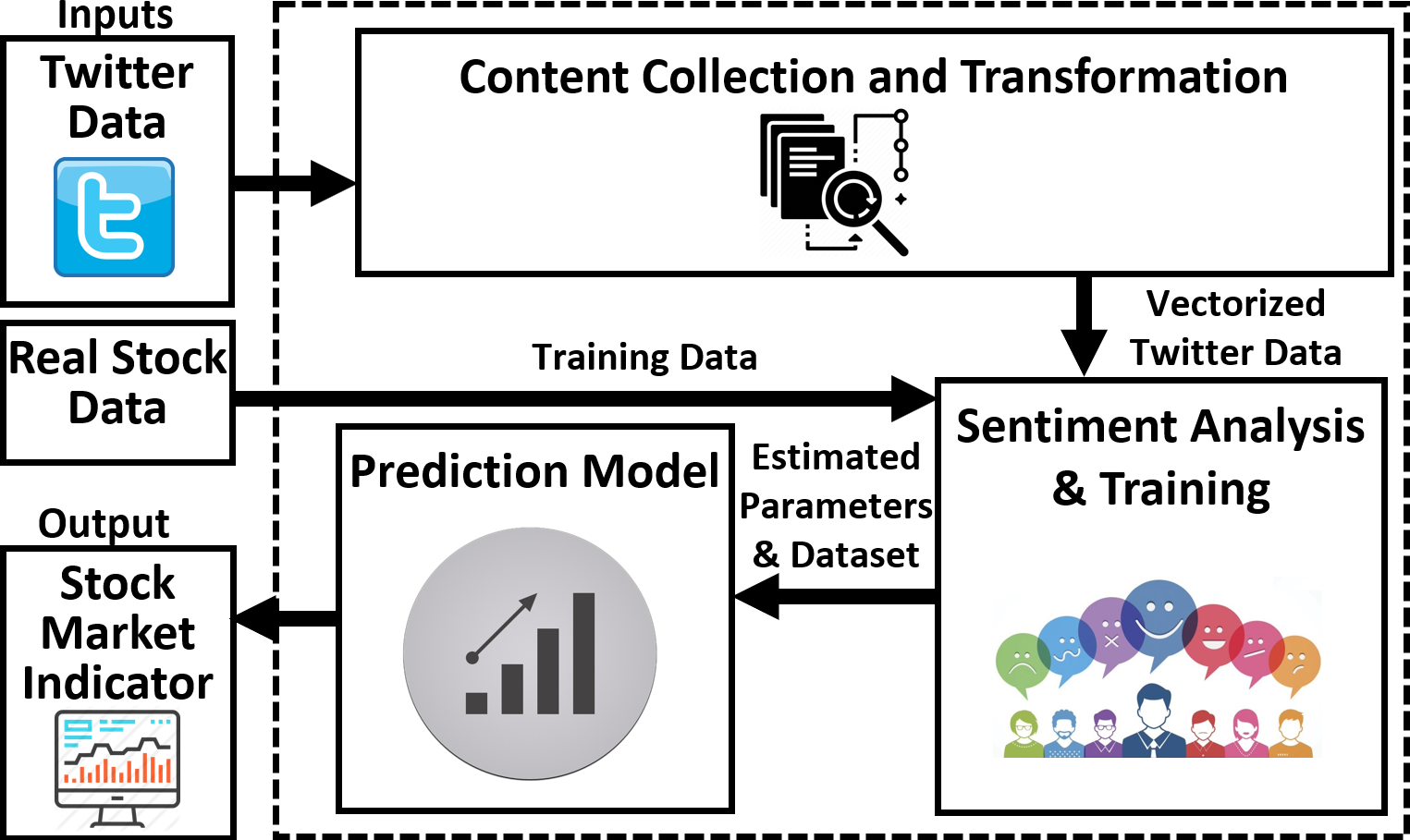}
    \caption{Block diagram of the Taureau Framework}
    \vspace{-0.15in}
    \label{fig:block}
\end{figure}

\subsection{Content Collection and Transformation}
In the first step, the framework collects, filters, organizes, and labels the noisy social media posts (i.e. tweets) to obtain an organized representation of the data for the analysis. Specifically, tweets from recent times fitting simple keyword search criteria are gathered. We obtained the tweets based on keyword searches using two Python based libraries: i) \textit{Tweepy}~\cite{roesslein2009tweepy} which can only collect real-time, and ii) \textit{getOldTweets}~\cite{henrique2017getoldtweets} which circumvents this restriction and can collect older tweets.

The collected data is subsequently filtered by running keyword searches on it (i.e. the names of the interested companies). The irrelevant data is discarded using \textit{regular expressions} and \textit{split strings}, and organized using a tokenizer called the \textit{nltk.tokenizer} package~\cite{perkins2010python}. Lastly the data is labeled into discrete opinions based on the nature of the tweets (e.g., praising a company's new product) using the \textit{nltk.classify} package~\cite{perkins2010python}.

\subsection{Sentiment Analysis and Training}
The Taureau framework leverages the TextBlob sentiment analysis engine~\cite{loria2018textblob} that uses an ensemble of different statistical approaches to generate the opinion and mood expression with the tweets. In particular, Taureau transforms the input texts using the Word2vec model~\cite{goldberg2014word2vec} and incorporates an opinion mining technique based on an ensemble of the Naive Bayes and Random Forest Classifiers to assess and quantify the mood of the users embedded within the tweets. 

For each tweet, the TextBlob engine generates two numeric outputs: i) the Subjectivity and ii) the Polarity. The Polarity, also known as orientation is the emotion expressed in the sentence~\cite{loria2018textblob}. The Polarity score ranges on a scale of $[-1,1]$ can be categorized as positive (i.e. closer to $1$), negative (i.e. closer to $-1$), or neutral (i.e. closer to $0$). Subjectivity, on the other hand is a measure of text as a fact or opinion. The metric is helpful in determining whether a text is an explanatory article based on facts (objective) or just an expression of a person's feeling or attitude about a company (subjective)~\cite{loria2018textblob}. The Subjectivity score ranges on a scale of $[0,1]$ with $0$ indicating least subjective and $1$ indicating most subjective tweets. Algorithm~\ref{alg:Subjectivity} summarizes on the process of obtaining these scores.

 \begin{algorithm} 
  	\footnotesize
  	\caption{Calculate Subjectivity and Polarity of each user}
  	\label{alg:Subjectivity} 
  	\hspace*{\algorithmicindent} \textbf{Input}: Tweets \\
  	\hspace*{\algorithmicindent} \textbf{Output}: Subjectivity = [], Polarity = []
  	\begin{algorithmic} [1]
  		\State {Initialize all Subjectivity and Polarity scores to $0$}
  		\While {index in ``length of cleansed\_dataframe"}
  		\State{wiki=TextBlob(``tweet\_text")}
  		\State{Subjectivity.append(wiki.sentiment.Subjectivity)}
  		\State{Polarity.append(wiki.sentiment.Polarity)}
  		\EndWhile
  		\State{cleansed\_dataframe[`Subjectivity]=Subjectivity}
  		\State{cleansed\_dataframe[`Polarity]=Polarity}
  	\end{algorithmic}
  \end{algorithm}
  
We use linear regression to connect the twitter sentiments with the actual historical stock data. To aggregate the data of the polarities for all the days, we use a series of steps based on~\cite{cocskun2018europehappinessmap}:

\begin{itemize}
    \item Find daily-sum polarities of each day ($\Sigma d+$, $\Sigma d-$)
    \item Count daily-number of tweets for each day ($\#d$)
    \item Find daily-average positive ($\mu d+$) and negative Polarity ($\mu d-$) of each day
    \item  Find meta-standard deviation of positive and negative polarities ($\sigma+$, $\sigma-$)
    \item Find meta-average of positive and negative polarities ($\mu+$, $\mu-$)
    \item Find daily sentiment Polarity for each day ($P_d$) 
\end{itemize} 

A similar approach is applied to aggregate the Subjectivity scores. Afterwards, we combine the Subjectivity and Polarity scores using a linear combinations having equal weights for both. The aggregate score is used for the stock inference process to predict the stock prices. The details of the stock inference process is described in the next subsection.
  
\subsection{Prediction Model}
For stock analysis, stock prices were converted to percent daily movement of adjusted stock price, following the example of the previous work in~\cite{bollen2011twitter}. Percent daily stock price movement was found to be strongly correlated between all companies and with the percent daily change in the Dow Jones Industrial Average. This trend was somewhat expected, especially since the companies under investigation are generally large and influential and likely to have some influence on, and to be influenced by, overall market performance. Further, a great deal of variability was observed in recent stock prices (since around early March 2020). This could probably be due to concerns surrounding the recent outbreak of the coronavirus disease 2019 (COVID-19). For these reasons, we analyzed a ``corrected" daily percent stock price movement, which is calculated by subtracting the Dow Jones' percent movement from the company stock's percent movement, with the goal of eliminating some of the noise from recent stock market volatility and allowing investigation of trends that are specific to the performance of each company.
 
Afterwards, we correlate the temporal dimensions of the obtained tweets with daily stock data to train and test our model to predict stock movement from lagged values of sentiment scores. Based on the methods of ~\cite{shah2018predicting}, Taureau outputs the continuous numerical value for predicted percent stock movement which is translated to a buy $(>0.5\%)$, hold $(-0.5\%,0.5\%)$, or sell $(<-0.5\%)$ recommendation. We also validate our model afterwards. For example, a predicted buy is correct if actual movement $>0.5\%$, predicted hold is correct if actual movement $(-1\%,1\%)$, and predicted sell is correct if actual movement $<0.5\%$.

Last but not the least, the Taureau framework will attempt to identify potential reasons for anticipated large changes in stock price. During the  period of March-April 2020, several companies experienced daily stock price changes of magnitude greater than 10\%, and Tesla experienced corrected daily stock price movements both greater than 10\% and less than -10\%. For any predicted movement of magnitude larger than 10\%\footnote{The threshold can be changed, but is set at 10\% by default and for demonstration.}, Taureau will find and report the 15\footnote{Again, the threshold is by default 15 but can be changed.} most repeated tweets from that day, as a snapshot of some important topics of conversation that may have led to that extreme prediction. This functionality will be validated manually by applying it to past data points with a change greater than 10\% and cross-referencing the reported claims about that company with any events reported in the news that may have driven that company's stock performance.

%% file: results.tex
\section{Evaluation \& Discussion} \label{sec:results}

\subsection{Task 1: Data Collection and Pre-Processing}

The first stepping stone is to gather the dataset for our framework. We collected real world datasets using Twitter data feeds for seven companies of interest: Apple, Google, Tesla, Facebook, Intel, T-mobile, and Amazon. As mentioned earlier, we obtained the tweets based on keyword searches using the two Python based libraries: i) \textit{Tweepy}~\cite{roesslein2009tweepy}, and ii) \textit{getOldTweets}~\cite{henrique2017getoldtweets}. The time period for the dataset is between March 6, 2020 and April 25, 2020. As highlighted above, a series of pre-processing steps are carried out to condition the input data into a desirable form.

\subsection{Task 2: Sentiment Analysis and Validation}

The key principle of sentiment analysis is to determine the Polarity of human emotions that can be broadly classified into happiness, affection, sorrow, grief, hatred, anger, and so on~\cite{gruhl2005predictive}. Opinion characterization is a domain of sentiment analysis where the sentiment of the text is extracted with respect to mostly online sources like social media. One challenge in opinion characterization during sentiment analysis is to handle the misspellings, colloquial language, short forms, and emoticons, all of which make it difficult to identify the appropriate sentiment of each tweet. Several domains of recent researches have made attempts to workaround this challenge using statistical analysis and machine learning concepts~\cite{gruhl2005predictive,apala2013prediction}. 

We used TextBlob to obtain the sentiment scores of the incoming tweets. In order to perform validation of the analysis, we subsequently compared the scores with our own observations to assess the accuracy of the generated sentiment score. In general, we found a good connection between the sentiment scores with our perception. We also found that regardless of the existence of misspellings, colloquial language, short forms, and emoticons, TextBlob performed a good job in characterizing the opinions with substantial accuracy. This implies that the scoring model used is trustworthy and can be applied to general use case scenarios.

\subsection{Task 3: Stock Movement Prediction}

Once the daily sentiment scores had been obtained, the first step of exploratory data analysis was to investigate a correlation between the corrected daily stock movement and the sentiment scores for that day. These results are shown in Table~\ref{tab:corr}. There is very weak correlation among measurements on the same day: the correlation between stock movement and Polarity is -0.044, and between stock movement and Subjectivity is 0.081, indicating almost non-existent correlation. The sentiment scores, like the tweets that generate them, are likely to include a lot of noise, so the same correlation analysis was done using sliding window averaging. Window sizes of 1, 3, 5, 7, and 9 were used as the number of surrounding days with which to average each sentiment score (days without this many neighbors were left unchanged).

\begin{table}[htb]
\scriptsize
    \caption{Correlation with Sliding Window Averaging}
\vspace{-0.1in}
    \centering
    \begin{tabular}{|l|l|l|l|l|l|}
        \hline
        Window Size & 1 & 3 & 5 & 7 & 9 \\ \hline
        Polarity Correlation & -0.044 & -0.345 & -0.321 & -0.313 & -0.197 \\ \hline
        Subjectivity Correlation & 0.081 & -0.225 & -0.155 & -0.134 & -0.051 \\ \hline
        Aggregate Score Correlation & 0.036 & -0.299 & -0.227 & -0.204 & -0.102 \\ \hline
    \end{tabular}
    \label{tab:corr}
    \vspace{-0.1in}
\end{table}

\begin{figure}[!htb]
    \centering
    %\vspace{-0.08in}
    \includegraphics[width=8cm]{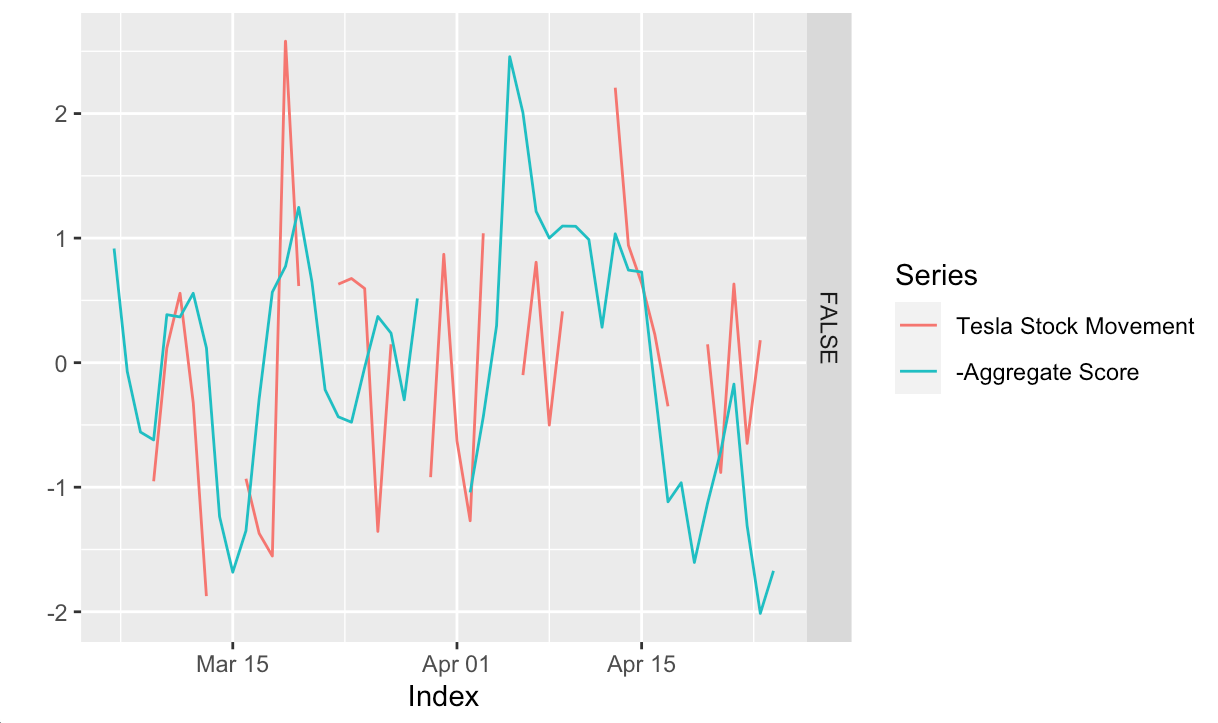}
    \vspace{-0.08in}
    \caption{Corrected daily stock movement against negative aggregate sentiment score (the average of Polarity and Subjectivity). Both measurements are standardized.}
    %\vspace{-0.08in}
    \label{fig:corr1}
    \vspace{-0.1in}
\end{figure}

\begin{figure}[!htb]
    \centering
    %\vspace{-0.08in}
    \includegraphics[width=8cm]{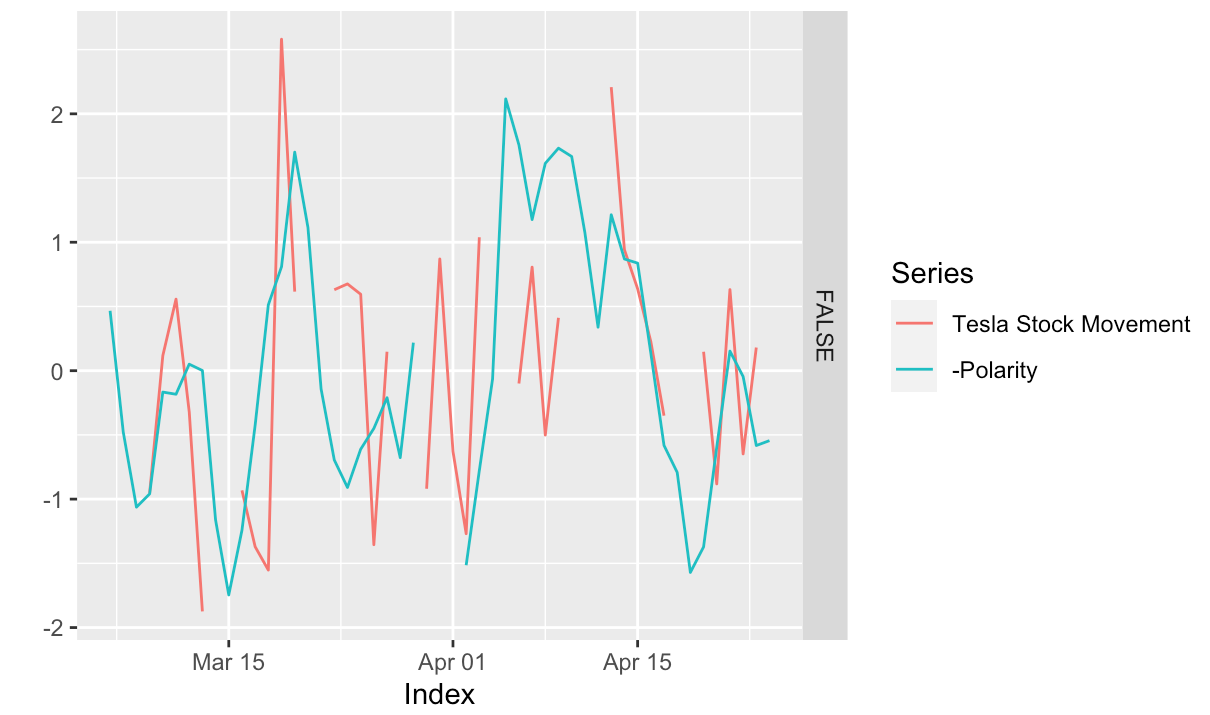}
    \vspace{-0.08in}
    \caption{Corrected daily stock movement against negative Polarity score. Both measurements are standardized.}
    %\vspace{-0.08in}
    \label{fig:corr2}
    \vspace{-0.1in}
\end{figure}

\begin{figure}[!htb]
    \centering
    %\vspace{-0.08in}
    \includegraphics[width=8cm]{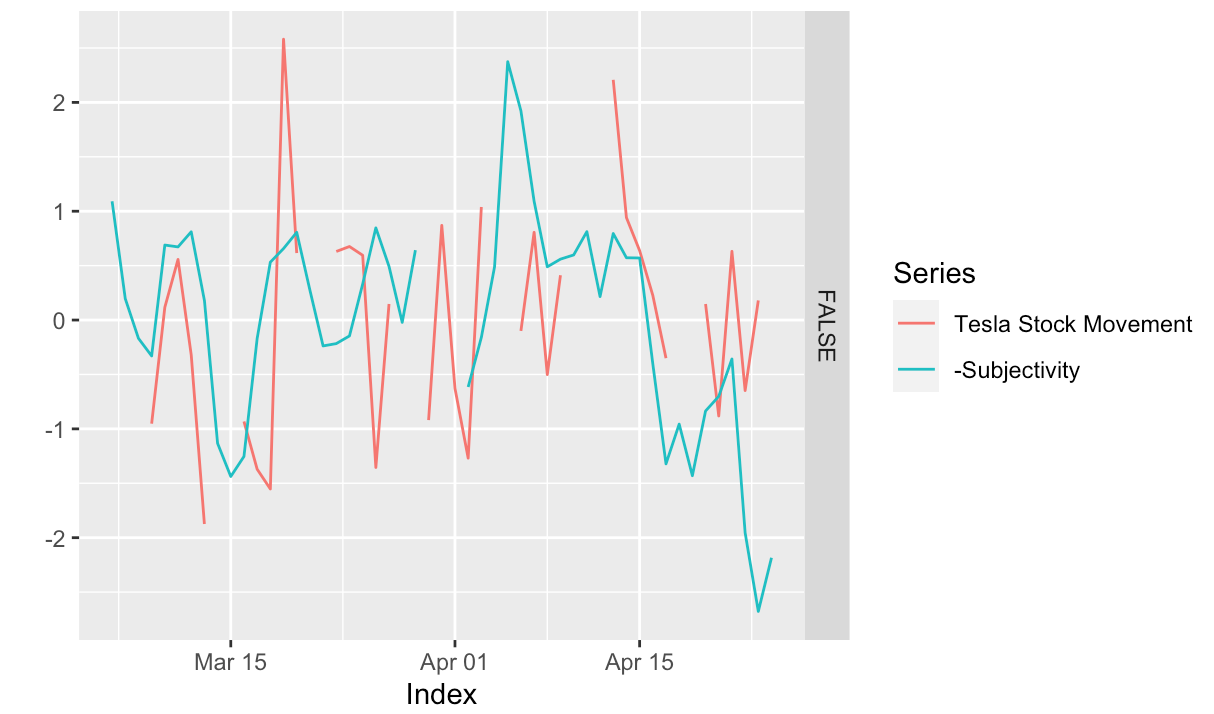}
    \vspace{-0.08in}
    \caption{Corrected daily stock movement against negative Subjectivity score. Both measurements are standardized.}
    %\vspace{-0.08in}
    \label{fig:corr3}
    \vspace{-0.1in}
\end{figure}

The correlation between stock movement and sentiment score with a window size of 3 is also shown in Figures~\ref{fig:corr1},~\ref{fig:corr2}, and~\ref{fig:corr3}. The negative of the sentiment scores is plotted due to the negative correlation. The correlation appears weak at best, but there does appear at least to be some directional correlation -- for example, stock movement and negative Polarity score tend to increase and decrease at many of the same points.

These results suggest that some averaging is helpful in filtering out noise; the strongest correlations are observed with a window size of 3. Using windows greater than 5 decreases the correlation, suggesting that windows that are too large introduce additional noise, which is also intuitive. However, these data also lead to the counter-intuitive conclusion that the sentiment scores are negatively correlated with stock movement. One would expect days on which the company's performance suffers to be dominated by negative sentiment on Twitter, and vice versa. Therefore, Polarity especially would be expected to positively correlated with stock movement, and it is unclear from investigating the data why the opposite is true. The possibility that correcting each stock's movement by the DJIA movement was causing this negative correlation was investigated, but this is not the cause because using uncorrected stock price movement gave similar conclusions and extremely similar correlation values. It was also observed that there is limited variability in the sentiment scores across days. The Polarity aspect of sentiment ranges from -1 to 1, but only values in the range [0.055, 0.226] were observed. Similarly, Subjectivity ranges from 0 to 1, but only values in the range [0.189, 0.471] were observed. From inspecting the data, the most common tweets tend to be dominated by positive tweets touting a new promotion, the success of a new product, etc. It may be that there is not as much capturable variability in the data as expected, but the correlation appears to be distinctly negative, and it is unclear why this is.

\begin{figure}[!htb]
    \centering
    %\vspace{-0.08in}
    \includegraphics[width=8cm]{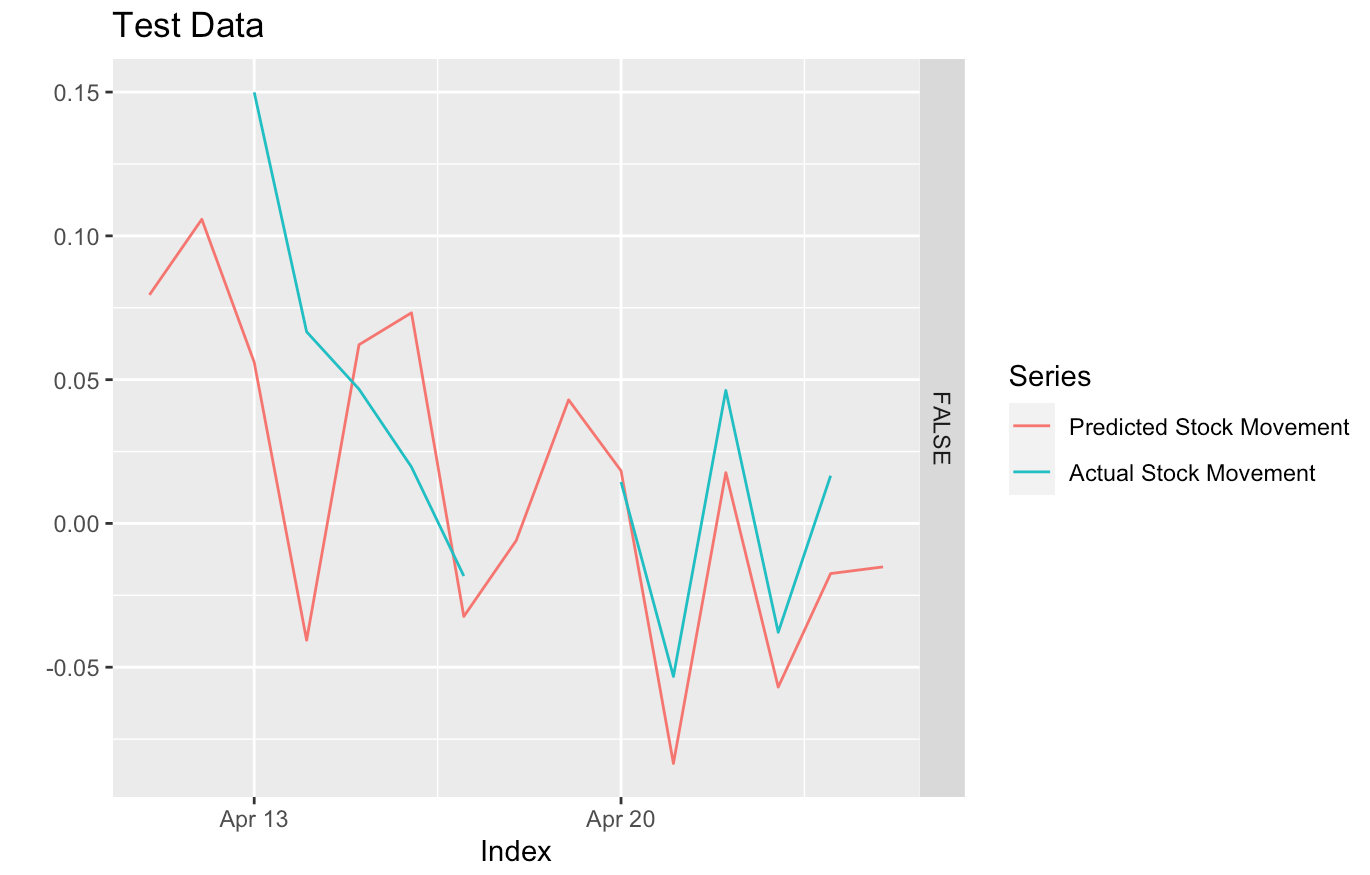}
    \caption{Predicted corrected stock price movement for the test dataset (April 11 - April 25) with the actual corrected movement}
    %\vspace{-0.08in}
    \label{fig:testdata}
    \vspace{-0.15in}
\end{figure}

To investigate the predictive potential of the data, a model was trained to predict corrected stock price movement from the Polarity and Subjectivity scores of 1, 2, and 3 days prior. The model was trained using data from March 6, 2020 to April 10, 2020 and tested on data from April 11, 2020 to April 25, 2020. The results of this prediction are shown in Figure~\ref{fig:testdata}. (Some values are missing because weekends are not trading days.) The output of this model is a predicted movement in the form of a decimal, which is converted to a recommendation (buy, hold, or sell) according to the description in Section~\ref{sec:proposed}.

The accuracy of this recommendation across the test data was 80\%. The ``null" accuracy for this method (using a model with random coefficients) is around 65\%. This result is encouraging, but should not be given too much weight due to the small sample size used (only 41 training observations and 10 test observations). More evidence of the inflating effect of the small sample size is the fact that the accuracy on the training data of 71\% is lower than the test set accuracy. Although not perfect, the model especially shows promise over the last five observations or so, during which time it follows the actual stock movement quite closely. It is unclear exactly why these days are more predictable, but it may be that there happens to be less volatility in the data over time as COVID-19 tensions begin to subside.

\subsection{Task 4: Tweet Reporting}

The final goal of the Taureau framework is to identify dates (past or future) with large predicted stock movements according to the predictive model of Task 3. To accomplish this, for any dates with a predicted movement of magnitude greater than some threshold (by default, 0.10, or 10\%), all the tweets about that company from that day are analyzed, and the 15 (by default) most-repeated tweets are reported, in an effort to identify any significant events from that time frame that may have contributed to that prediction.

It is difficult to analyze the usefulness of this reporting quantitatively, but in our analysis, one date (April 12, 2020) had a predicted movement greater than the threshold (+0.106, or +10.6\%). The most repeated tweet from this day was about the large number of orders for the Tesla Model 3, indicating the success of a new factory. Several were somewhat irrelevant conversations that happened to gain several retweets, but several more were predictions of future success for Tesla.

%% file: further.tex
\section{Conclusion \& Future Work} \label{sec:further}

\subsection{Challenges Faced}
The first challenge encountered involves data collection. Our aim was to collect tweets fitting a simple search criteria for seven large tech companies from January 1, 2019 to the present. We have been using the two tools \textit{Tweepy} and \textit{GetOldTweets}. Both tools are subject to the Twitter API's stringent rate limit, which made our data collection process harder than necessary. Moreover, getOldTweets occasionally crashed during operation. We were eventually able to overcome these limitations by incorporating delays for the data collection for every 15 minutes and building a helper program to accommodate program crashes. However, GetOldTweets also has the undesirable property of being quite slow as well. For these reasons, we were able to obtain complete data for one company (Tesla) for the period allowed with Tweepy (March 6, 2020 to present), but only intermittent data for previous dates.

Another major challenge faced, common to most social sensing projects, is the problem of ``noisy" social signals. Simple keyword searches on twitter often return unwanted information; for example, a keyword search for ``Tesla" returns results about the scientist Nikola Tesla rather than the company. Aside from manually sorting through thousands of tweets, for future work it might be effective to investigate constructing a classifier to categorize tweets as about that company or not. Similar to this is the issue of subjectivity. People tend to be sarcastic, skeptical, and emotional, especially during market volatility as is going on currently. Such subjective tweets might impart bias into the computation of the sentiments. The subjectivity facet of sentiment scores was included in an effort to capture and correct for this.

Significant challenges were also anticipated in any current effort to analyze the stock market due to the prevalence of the coronavirus. Due to the extreme concerns about the market and the ensuing crash, it was expected to be difficult to extract information about performance that is specific to each company. Variability in stock prices increased greatly beginning on March 12, around the time major events such as sporting events began to be canceled and the coronavirus was brought to the front of public attention. It was our goal that correcting each movement by the Dow Jones movement will at least partially correct for this.

One final big challenge encountered was the lack of up-to-date training corpus for sentiment analysis. In order to perform sentiment analysis on a new set of data with satisfactory accuracy, a recent prior training corpus of text is necessary to train the sentiment generation model by determining its parameter values~\cite{pak2010twitter}. In TextBlob, a set of training and testing data can be specified which is used to train and validate the sentiment analysis process~\cite{loria2018textblob}. While general sentiment analysis models yield accurate sentiment analysis on Twitter data, our case is quite different due to presence of a significant amount of additional noise in the twitter data caused by the ongoing COVID-19 pandemic. With the highly volatile public attitude and stock market in recent times, training corpuses get old very fast. A corpus that seemingly provided desirable sentiment scores with tweets generated weeks ago may not be good enough for new data generated in a present week. Thus, we found it difficult to train sentiment analysis with limited number of available corpuses. To work around this issue, we tested various corpuses from multiple sources (similar to ensemble training) and obtained  the sentiment scores. We performed cross-validation and retained the one that gave the best performance.

\subsection{Future Directions}

The most notable limitation of this study was the small sample size available. Due to the difficulties associated with data collection, cleaning, and processing, only 51 data points were available for analysis. This severely limited the scope of conclusions that were able to be made: because of the small sample size, it is difficult to validate the model, so it is unclear whether the observed accuracy truly identifies a pattern or is due to chance. In the final predictive model, the coefficients associated with polarity and subjectivity are positive at certain lags and negative at others, suggesting that the model did not capture true patterns but rather noise and chance. It is difficult to manually identify patterns either, as it is impossible to observe long-term trends in public opinion and examine their correlation to the stock market. Compounding this problem is the fact that the time period for which data was collected, March-April 2020, was an extremely volatile time both for the stock market and for public mood. Both signals are likely to have an exceedingly large amount of noise during this time period, and even if trends are able to be identified through this, the period's uniqueness means it is dubious whether those trends would generalize to ``normal," less volatile time periods. Even worse, Tesla occupies a unique position in the public mood, as CEO Elon Musk is a captivating and somewhat controversial figure, so it is unclear whether any conclusions drawn about Tesla could generalize to other companies. It was originally a goal to investigate extending our model first to similar companies (large tech companies with easy brand recognition), and eventually to companies in other industries (e.g. fashion). Increasing the number of available data points would make it possible to identify whether any patterns do emerge between stock movement and sentiment scores. After identifying any such patterns, it might be possible to capture them with a more complex model, such as one that models interactions between the different facets of sentiment scores, or one that handles non-linearity in the data like a neural network.

This is an inherently interdisciplinary problem. The authors together have background in social sensing methodology and mathematical modeling techniques, but have limited knowledge of financial markets. Having a more complete understanding of the application of this problem would inevitably allow for a more accurate and informed analysis of public sentiment as it relates to company performance, and a more meaningful interpretation of the results. For example, it seems as though both polarity and subjectivity correlate negatively with stock movement, which seems counter-intuitive, but may in fact have a simple explanation to a subject matter expert. Further, the ultimate application of this analysis would be to use the techniques developed in Taureau as an automated stock trading engine. It would therefore be informative to investigate whether the final predictive model is capable of making informed trading decisions, for example, by simulating a few weeks of stock market trading and determining whether Taureau makes or loses money.

From this discussion, another obvious extension of the Taureau framework would be to develop functionality to identify new companies that are predicted to perform well in which to invest. This might be done by extending existing topic-sentiment modeling to identify topics about which public mood is overwhelmingly positive, and developing techniques to identify whether there is a company associated with that topic that might be worth investing in.

%% file: main.bbl
\begin{thebibliography}{10}
\providecommand{\url}[1]{#1}
\csname url@samestyle\endcsname
\providecommand{\newblock}{\relax}
\providecommand{\bibinfo}[2]{#2}
\providecommand{\BIBentrySTDinterwordspacing}{\spaceskip=0pt\relax}
\providecommand{\BIBentryALTinterwordstretchfactor}{4}
\providecommand{\BIBentryALTinterwordspacing}{\spaceskip=\fontdimen2\font plus
\BIBentryALTinterwordstretchfactor\fontdimen3\font minus
  \fontdimen4\font\relax}
\providecommand{\BIBforeignlanguage}[2]{{%
\expandafter\ifx\csname l@#1\endcsname\relax
\typeout{** WARNING: IEEEtran.bst: No hyphenation pattern has been}%
\typeout{** loaded for the language `#1'. Using the pattern for}%
\typeout{** the default language instead.}%
\else
\language=\csname l@#1\endcsname
\fi
#2}}
\providecommand{\BIBdecl}{\relax}
\BIBdecl

\bibitem{zhang2020transres}
Y.~Zhang, R.~Zong, J.~Han, D.~Zhang, T.~Rashid, and D.~Wang, ``Transres: a deep
  transfer learning approach to migratable image super-resolution in remote
  urban sensing,'' in \emph{2020 17th annual IEEE international conference on
  sensing, communication, and networking (SECON)}.\hskip 1em plus 0.5em minus
  0.4em\relax IEEE, 2020, pp. 1--9.

\bibitem{zhang2020pqa}
Y.~Zhang, X.~Dong, M.~T. Rashid, L.~Shang, J.~Han, D.~Zhang, and D.~Wang,
  ``Pqa-cnn: Towards perceptual quality assured single-image super-resolution
  in remote sensing,'' in \emph{2020 IEEE/ACM 28th International Symposium on
  Quality of Service (IWQoS)}.\hskip 1em plus 0.5em minus 0.4em\relax IEEE,
  2020, pp. 1--10.

\bibitem{bollen2011twitter}
J.~Bollen, H.~Mao, and X.~Zeng, ``Twitter mood predicts the stock market,''
  \emph{Journal of computational science}, vol.~2, no.~1, pp. 1--8, 2011.

\bibitem{nguyen2015sentiment}
T.~H. Nguyen, K.~Shirai, and J.~Velcin, ``Sentiment analysis on social media
  for stock movement prediction,'' \emph{Expert Systems with Applications},
  vol.~42, no.~24, pp. 9603--9611, 2015.

\bibitem{shah2018predicting}
D.~Shah, H.~Isah, and F.~Zulkernine, ``Predicting the effects of news
  sentiments on the stock market,'' in \emph{2018 IEEE International Conference
  on Big Data (Big Data)}.\hskip 1em plus 0.5em minus 0.4em\relax IEEE, 2018,
  pp. 4705--4708.

\bibitem{loria2018textblob}
S.~Loria, ``textblob documentation,'' \emph{Release 0.15}, vol.~2, 2018.

\bibitem{hasan2016development}
S.~M. Hasan, M.~T. Rashid, M.~S.~S. Chowdhury, and M.~K. Rhaman, ``Development
  of a credible and integrated electronic voting machine based on contactless
  ic cards, biometrie fingerprint credentials and pos printer,'' in \emph{2016
  IEEE Canadian Conference on Electrical and Computer Engineering
  (CCECE)}.\hskip 1em plus 0.5em minus 0.4em\relax IEEE, 2016, pp. 1--5.

\bibitem{wang2018age}
D.~Wang, B.~K. Szymanski, T.~Abdelzaher, H.~Ji, and L.~Kaplan, ``The age of
  social sensing,'' \emph{IEEE Computer}, 2018.

\bibitem{liu2018urban}
Z.~Liu, Z.~Li, K.~Wu, and M.~Li, ``Urban traffic prediction from mobility data
  using deep learning,'' \emph{IEEE Network}, vol.~32, no.~4, pp. 40--46, 2018.

\bibitem{mao2012citysee}
X.~Mao, X.~Miao, Y.~He, X.-Y. Li, and Y.~Liu, ``Citysee: Urban co 2 monitoring
  with sensors,'' in \emph{INFOCOM, 2012 Proceedings IEEE}.\hskip 1em plus
  0.5em minus 0.4em\relax IEEE, 2012, pp. 1611--1619.

\bibitem{rashid2019socialcar}
M.~T. Rashid, D.~Y. Zhang, and D.~Wang, ``Socialcar: A task allocation
  framework for social media driven vehicular network sensing systems,'' in
  \emph{2019 15th International Conference on Mobile Ad-Hoc and Sensor Networks
  (MSN)}.\hskip 1em plus 0.5em minus 0.4em\relax IEEE, 2019, pp. 125--130.

\bibitem{rashid2020dasc}
------, ``Dasc: Towards a road damage-aware social-media-driven car sensing
  framework for disaster response applications,'' \emph{Pervasive and Mobile
  Computing}, vol.~67, p. 101207, 2020.

\bibitem{gruhl2005predictive}
D.~Gruhl, R.~Guha, R.~Kumar, J.~Novak, and A.~Tomkins, ``The predictive power
  of online chatter,'' in \emph{Proceedings of the eleventh ACM SIGKDD
  international conference on Knowledge discovery in data mining}, 2005, pp.
  78--87.

\bibitem{apala2013prediction}
K.~R. Apala, M.~Jose, S.~Motnam, C.-C. Chan, K.~J. Liszka, and F.~de~Gregorio,
  ``Prediction of movies box office performance using social media,'' in
  \emph{2013 IEEE/ACM International Conference on Advances in Social Networks
  Analysis and Mining (ASONAM 2013)}.\hskip 1em plus 0.5em minus 0.4em\relax
  IEEE, 2013, pp. 1209--1214.

\bibitem{mittal2012stock}
A.~Mittal and A.~Goel, ``Stock prediction using twitter sentiment analysis,''
  \emph{Standford University, CS229 (2011 http://cs229. stanford.
  edu/proj2011/GoelMittal-StockMarketPredictionUsingTwitterSentimentAnalysis.
  pdf)}, vol.~15, 2012.

\bibitem{medhat2014sentiment}
W.~Medhat, A.~Hassan, and H.~Korashy, ``Sentiment analysis algorithms and
  applications: A survey,'' \emph{Ain Shams engineering journal}, vol.~5,
  no.~4, pp. 1093--1113, 2014.

\bibitem{rao2012analyzing}
T.~Rao, S.~Srivastava \emph{et~al.}, ``Analyzing stock market movements using
  twitter sentiment analysis,'' \emph{ASONAM 2012}, 2012.

\bibitem{roesslein2009tweepy}
J.~Roesslein, ``tweepy documentation,'' \emph{Online] http://tweepy.
  readthedocs. io/en/v3}, vol.~5, 2009.

\bibitem{henrique2017getoldtweets}
J.~Henrique, ``Getoldtweets-python,'' \emph{Jefferson-Henrique}, 2017.

\bibitem{perkins2010python}
J.~Perkins, \emph{Python text processing with NLTK 2.0 cookbook}.\hskip 1em
  plus 0.5em minus 0.4em\relax Packt Publishing Ltd, 2010.

\bibitem{goldberg2014word2vec}
Y.~Goldberg and O.~Levy, ``word2vec explained: deriving mikolov et al.'s
  negative-sampling word-embedding method,'' \emph{arXiv preprint
  arXiv:1402.3722}, 2014.

\bibitem{cocskun2018europehappinessmap}
M.~Co{\c{s}}kun and M.~Ozturan, ``\# europehappinessmap: A framework for
  multi-lingual sentiment analysis via social media big data (a twitter case
  study),'' \emph{Information}, vol.~9, no.~5, p. 102, 2018.

\bibitem{pak2010twitter}
A.~Pak and P.~Paroubek, ``Twitter as a corpus for sentiment analysis and
  opinion mining.'' in \emph{LREc}, vol.~10, no. 2010, 2010, pp. 1320--1326.

\end{thebibliography}
